\documentclass[]{aa}
\usepackage{amssymb,amsmath}
\usepackage{graphicx}
\usepackage{xcolor,natbib}
\usepackage{multirow}
\usepackage[T1]{fontenc}
\usepackage{ae,aecompl}
\usepackage{newtxtext, newtxmath}
\usepackage{float}
\usepackage{subfigure}
\usepackage{longtable}
\usepackage{enumitem}
\usepackage{longtable,listings}
\usepackage[flushleft]{threeparttable}
\usepackage{parcolumns}
\usepackage{hyperref}

\begin{document}

\title{Evolution of broad emission lines from double-peaked to single-peaked to support a central tidal disruption event}

%\correspondingauthor{XueGuang Zhang}% \email{xgzhang@njnu.edu.cn}}
%\email{xgzhang@gxu.edu.cn}
%\author{XueGuang Zhang$^{*}$}
%\affiliation{Guangxi Key Laboratory for Relativistic Astrophysics, School of Physical Science and Technology, 
%GuangXi University, No. 100, Daxue Road, Nanning, 530004, P. R. China}

\titlerunning{unique evolution of double-peaked broad lines in TDEs}

\author{XueGuang Zhang}
\institute{Guangxi Key Laboratory for Relativistic Astrophysics, School of Physical Science and Technology,
GuangXi University, Nanning, 530004, P. R. China \ \ \ \ \email{xgzhang@gxu.edu.cn}}

\abstract{ %%%about 192 words
In this manuscript, considering evolution of fallback accreting debris in a central Tidal Disruption Event (TDE), the 
outer boundary increased with time of the disk-like broad emission line regions (BLRs) lying into central accretion disk will 
lead expected broad emission lines changed from double-peaked to single-peaked. Considering common elliptical orbitals for the 
accreting fallback TDEs debris, based on simulated results through the preferred standard elliptical accretion disk model, a 
probability about 3.95\% can be estimated for cases with double-peaked profile changed to single-peaked profile in 
multi-epoch broad emission lines, indicating such unique profile variability could be indicator for BLRs related to TDE debris. 
Meanwhile, among the reported optical TDE candidates with apparent broad lines, such profile changes in broad H$\alpha$ can be 
found in the AT 2018hyz. After accepted the outer boundaries of the disk-like BLRs increased with time, the observed multi-epoch 
broad H$\alpha$ can be described in AT 2018hyz. Moreover, the elliptical accretion disk model determined time dependent ratios of 
the outer boundaries of the disk-like BLRs are well consistent with the TDE model expected ratios of the outer boundaries of the 
fallback TDE debris. Furthermore, the evolution properties of disk-like BLRs can be applied to estimate the locations of the 
disk-like BLRs of which outer boundary could be about one sixth of the outer boundary of the fallback TDE debris in AT 2018hyz. 
Such unique profile changes from double-peaked to single-peaked could be applied as further clues to support a central TDE.
}


\maketitle

\section{Introduction}
 
%%%1
	Broad emission lines can be detected in optical spectrum in a Tidal Disruption Event (TDE), due to expected broad emission 
line regions (BLRs) lying into central accretion disk related to fallback TDE debris, as detailed discussions in \citet{gm14}. 
Moreover, such disk-like BLRs lying into central accretion disk can commonly lead to double-peaked broad emission lines, as detailed 
discussions on accretion disk model in \citet{ch89a, ch89b, el95, se03, fe08}, etc. The model expected double-peaked broad emission 
lines have been reported in more than 400 broad line Active Galactic Nuclei (AGN), as reported results in \citet{eh94, st03, zh22a, 
zh23, wg24, zh24b}, etc. Meanwhile, there are some reports on double-peaked broad emission lines in TDE candidates, such as the 
discussed results in \citet{yw13, lz17, hf20, sn20, zh21, sl23, rn24, zh24a}, etc. However, there are different causes leading to 
variability of the double-peaked broad emission lines in AGN and in TDE candidates.

%%%2
	The unique disk-like structures of BLRs lying into central accretion disks can lead to unique time-dependent variability 
properties of the corresponding broad emission lines, commonly due to rotating motions in the disk-like BLRs, especially in broad 
line AGN. \citet{se03, ss12} have shown such variability of the double-peaked broad H$\alpha$ in the known NGC1097 by rotating 
disk-like BLRs lying into central accretion disk. Similar variability properties of double-peaked broad emission lines can also 
be found in \citet{ge07} and in \citet{le10} in small samples of broad line AGN.

\begin{figure*}
\centering\includegraphics[width = 18cm,height=8cm]{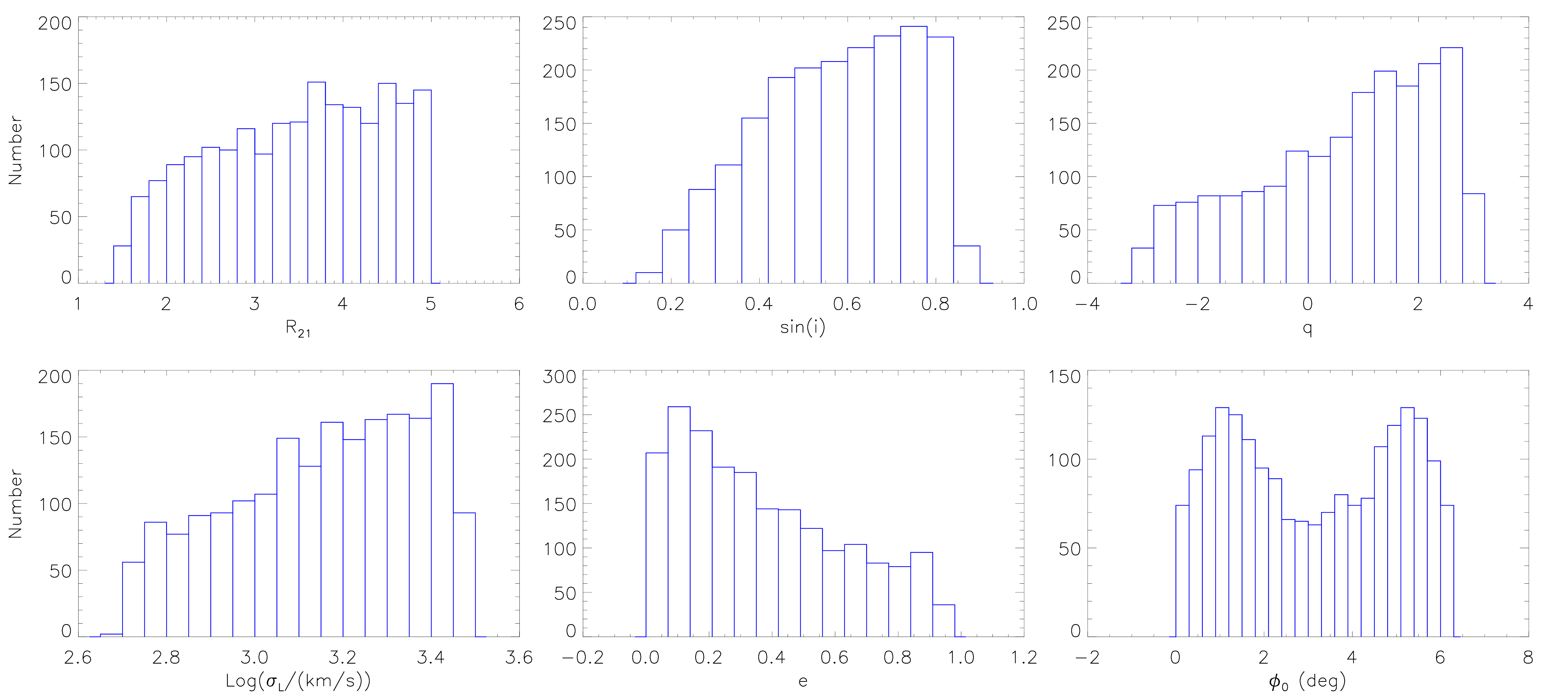}
\caption{Distributions of the model parameters of the 1977 cases which have profiles changed from double-peaked to single-leaked.}
\label{dis}
\end{figure*}

%%%3
	Although there are similar accretion disk origin for the reported double-peaked broad emission lines in normal broad line 
AGN and in TDE candidates, one main difference should be noted between the physical properties of the corresponding BLRs in normal 
broad line AGN and in TDE candidates. As the discussed TDE candidates in \citet{gs21} and the more recent reported TDE candidates 
in \citet{yr23}, the time durations of TDE candidates are around 1yr (mean values for the cases shown in Fig.~9 in \citealt{yr23}), 
very smaller than the variability time scale in double-peaked broad lines in AGN as shown in \citet{se03, ge07, le10}. The shorter 
time durations of TDE candidates probably indicate few effects of rotations in disk-like BLRs in TDE candidates on the corresponding 
broad lines. However, when considering the BLRs related TDE debris, the boundaries of the disk-like BLRs should increase with time, 
due to the time dependent evolution of fallback TDE debris, such as the shown results in \citet{gm14} for the outer 
boundary of TDE debris after considering the dimensionless evolution time $t = T/T_0$ with $T_0 =2\pi(\frac{R_p^3}{G~M_{BH}})^{1/2}$ 
($R_p$ as the pericenter distance) and $T$ as the orbital period at radius different from $R_p$ (such as in \citealt{lo11})
\begin{equation}
	R_{tde, out} \sim 2(\frac{G~M_{BH} t^2}{\pi^2})^{1/3} 
\end{equation}
with $M_{BH}$ and $G$ as the central black hole (BH) mass and the Gravitational constant. Therefore, not similar as the commonly 
known disk-like BLRs in normal broad line AGN, the variability in line profiles of the broad lines from the disk-like BLRs in 
TDE candidates is not due to rotations, but probably mainly due to the variability of the outer boundary of the disk-like BLRs. 
To check the effects of the time dependent variability of outer boundary of the disk-like BLRs in TDE candidates is the main 
objective of this manuscript. Here, we should note that starting from evolution of outer radius of disk-like BLRs in 
TDEs, line profile variability of double-peaked broad lines are mainly considered in our manuscript, not to determine physical 
origin of line profile variability of double-peaked broad lines.

%%%4
	This manuscript is organized as follows. Section 2 presents our main results through the elliptical accretion disk model 
\citep{el95} and necessary discussions on variability properties of double-peaked broad lines due to variability of outer boundary 
of the disk-like BLRs in TDE candidates. Section 3 shows discussions on another explanations to the changed profiles of broad 
emission lines. Section 4 gives our main summary and conclusions. And in this manuscript, we have adopted the cosmological 
parameters of $H_{0}=70{\rm km\cdot s}^{-1}{\rm Mpc}^{-1}$, $\Omega_{\Lambda}=0.7$ and $\Omega_{\rm m}=0.3$.

\section{Main results}

\begin{table}
\caption{Model parameters for the elliptical accretion disk model}
\begin{tabular}{llll}
\hline\hline
        par & units & model & Elli  \\
\hline\hline
	$R_{in}$ & $\rm R_G$ & [100,~600]    &  166$\pm$90\\
	$R_{out}$ & $\rm R_G$ & $k_0\times R_{in}$ ($k_0\in[1.5,~5]$) & 880$\pm$113 \\
	$R_{out2}$ & $\rm R_G$ & $k\times R_{out}$ ($k\in[1.5,~5]$) &  1006$\pm$260 \\
	           & $\rm R_G$ &                                     &  1236$\pm$370  \\
		   & $\rm R_G$ &                                     &  1340$\pm$502  \\
		   & $\rm R_G$ &                                     &  1900$\pm$764  \\
	$\sin(i)$ &   & [0.3, ~0.9]   & 0.62$\pm$0.03 \\
        $q$ &         & [-3, ~3]  &   -0.103$\pm$0.687\\
        $e$ &         & [0, ~0.95]  &  0.133$\pm$0.018 \\
        $\sigma_L$ & km/s & [500, ~3000]  &  1893$\pm$200\\
	&     km/s    &            &     750$\pm$164 \\
	&     km/s    &            &     1720$\pm$220 \\
	&     km/s    &            &     1818$\pm$198 \\
	&     km/s    &            &     2515$\pm$357 \\
	$\phi_0$      & rad  & [0, ~2$\pi$]  &  0.12$\pm$0.04\\
\hline\hline
\end{tabular}\\
Notice: The first column and the second column show the applied model parameters and the corresponding units. The third column shows 
the limited range ([lower boundary, upper boundary]) of each model parameter in the elliptical accretion disk model leading to the 
artificial $f_{\lambda1}$ and $f_{\lambda2}$. The fourth column shows the determined model parameters in the elliptical accretion disk 
model applied to describe the time dependent line profiles of broad H$\alpha$ in AT 2018hyz. In the fourth column, the five values 
in $R_{out}$ and $R_{out2}$ and the five values in $\sigma_L$ are for the five broad H$\alpha$ with $\Delta t=17, 51, 117, 120, 164$days.
\end{table}

%%%2.1
	In this manuscript, the elliptical accretion disk model proposed in \citet{el95} is accepted to describe expected double-peaked 
broad emission lines from disk-like BLRs lying into central accretion disk. There are seven free model parameters, the inner boundary 
$R_{in}$ (in units of $R_{G}$ with $R_G$ as the Schwarzschild radius) and the outer boundary $R_{out}$ (in units of $R_{G}$) of the 
disk-like emission regions, the eccentricity $e$ of the emission regions, the inclination angle $i$ of the emission regions, the line 
emissivity power-law index $f_r\propto r^{-q}$, the local turbulent broadening velocity $\sigma_L$ (in units of km/s), and the 
orientation angle $\phi_0$. Here, the improved circular accretion disk plus spiral arms model \citep{se03} is not considered, mainly 
due to the following reason. As discussed in \citet{gm14}, the fallback TDE debris are commonly in elliptical orbits, such as the shown 
results in Fig.~2 in \citet{gm14}, therefore, the standard elliptical accretion disk model discussed in \citet{el95} is mainly considered. 
Simple discussions on the circular accretion disk plus spiral arms model \citep{se03} should be given in the following Section 3.

%%%2.2
	Based on the elliptical accretion disk model, the first thing we should do is to estimate a probability to detect the profile 
change from double-peaked to single-peaked due to increased outer boundary $R_{out}$ with time. The following procedures are applied. 
For the first step, the seven model parameters are randomly collected within the limited ranges listed in Table~1, leading to a model 
created line profile $f_{\lambda1}$. Here, the model parameter $R_{out}$ is collected by $k_0\times R_{in}$ with $k_0$ as a random 
value from 1.5 to 5. And the listed range for each model parameter is common, see results in \citet{st03}. For the second step, a new 
value of the outer boundary $R_{out2}$ is randomly collected by $R_{out2}\sim k\times R_{out}$ ($k$ as a random value from 1.5 to 5) 
with $R_{out}$ as the value collected in the first step, leading to the new model created line profile $f_{\lambda2}$ with different 
outer boundaries but with the same values for the other model parameters. For the third step, if $f_{\lambda1}$ has more than two peaks 
but $f_{\lambda2}$ has only one peak, we accept the case is the one which has the changed profile from double-peaked to single-peaked. 
Here, the function find\_peaks (included in the package idlspec2d with version 5.2.0 in Sloan Digital Sky Survey) is applied to detect 
peaks in line profiles. Then, to repeat the three above steps 50000 times, there are 1977 cases having the changed profiles from the 
double-peaked to single-peaked. Although the model created cases are oversimplified, the results indicate the probability roughly about 
3.95\% (1977/50000) to detect changed profiles in multi-epoch optical spectrum, assumed that the outer boundaries of the disk-like BLRs 
in TDE candidates are increased with time. Moreover, based on the dynamical properties of the emission clouds in disk-like 
BLRs, the full width at zero intensity of the broad emission line sensitively depends on the inner boundary of the disk-like BLRs, but 
the peak separation of the broad emission line sensitively depends on the outer boundary. Larger outer boundary leads to smaller peak 
separation of the double-peaked broad emission line. Once the peak separation is smaller enough, single-peaked rather than apparent 
double-peaked broad emission line could be expected from disk-like BLRs.

	Here, we should note that our main objective to do the simulation is to check a probability for TDE expected evolution 
of outer radius of disk-like BLRs leading double-peaked broad emission line to be changed to single-peaked. Now, based on the randomly 
collected model parameters within reasonable limited ranges, a not very low probability about 3\%-4\% can be obtained, indicating 
among the reported more than 300 optical TDEs, there should be several TDEs which could have changed line profiles in their broad 
emission lines as expected by the simulations. Therefore, it is the basic point to support us in doing the following work in the 
manuscript. Unfortunately, through randomly collected model parameters in reasonable parameter space vast enough, if the corresponding 
simulations can not lead to an acceptable probability, it is very hard at current stage for us to find a more optimized way to 
determine a more trustworthy probability.

\begin{figure}
\centering\includegraphics[width = 8cm,height=5cm]{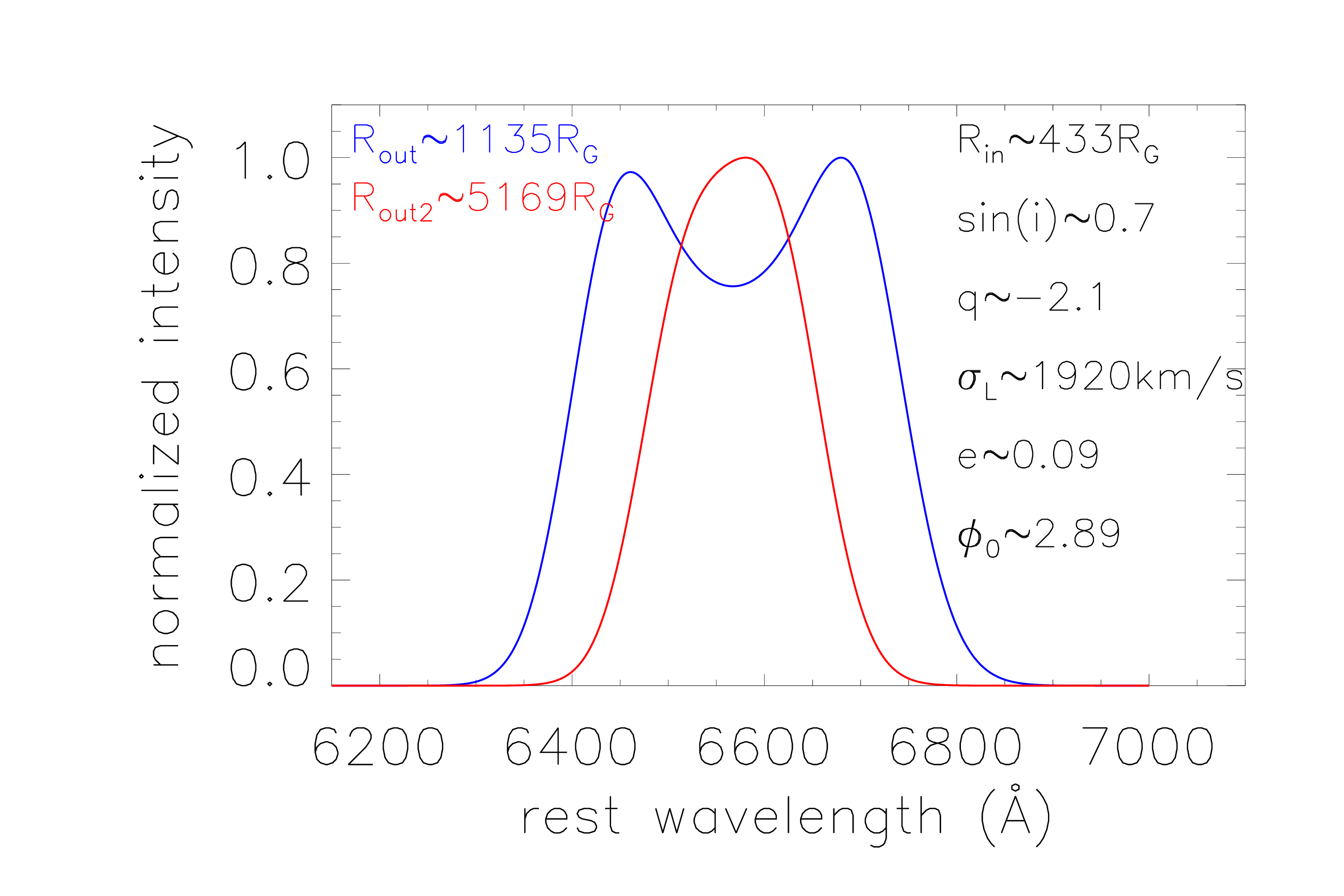}
\caption{An example on the broad H$\alpha$ with double-peaked profile (in blue) expected by the elliptical accretion disk model with 
smaller $R_{out}$, but with single-peaked profile (in red) expected by the model with larger $R_{out2}$. The applied model parameters, 
except the $R_{out}$ and $R_{out2}$ marked in the top left corner, are marked in the top right corner.}
\label{mod}
\end{figure}

\begin{figure*}
\centering\includegraphics[width = 18cm,height=12cm]{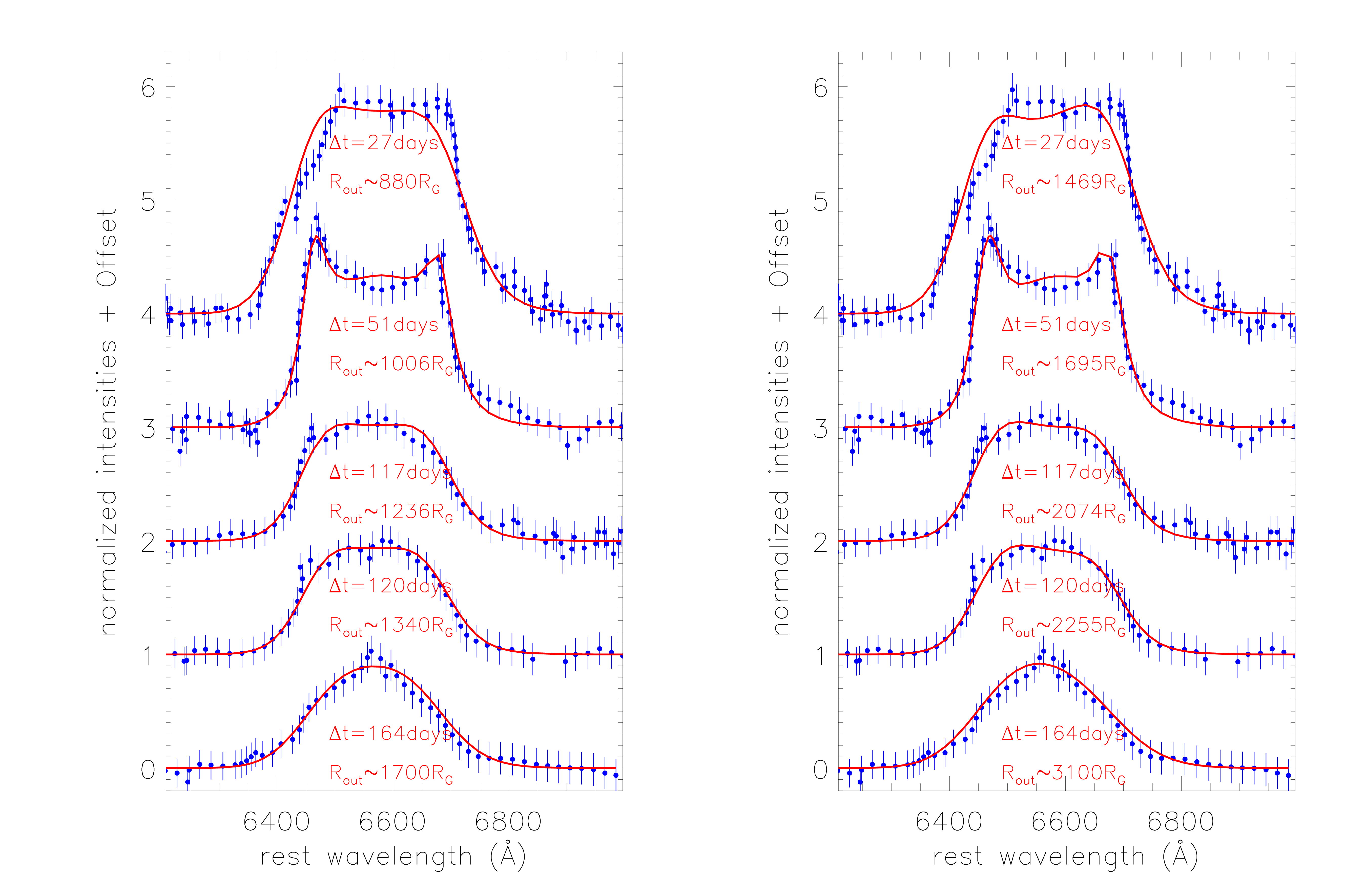}
\caption{Left panel shows the best descriptions (solid red lines) to the multi-epoch broad H$\alpha$ (solid circles plus error bars 
in blue) in AT 2018hyz by the elliptical accretion disk model. The corresponding $\Delta t$ and outer boundary of the disk-like BLRs 
for each broad H$\alpha$ are marked as red characters. Right panel shows the corresponding results through the circular 
accretion disk plus spiral arms model.}
\label{tde}
\end{figure*}

\begin{figure*}
\centering\includegraphics[width = 18cm,height=12cm]{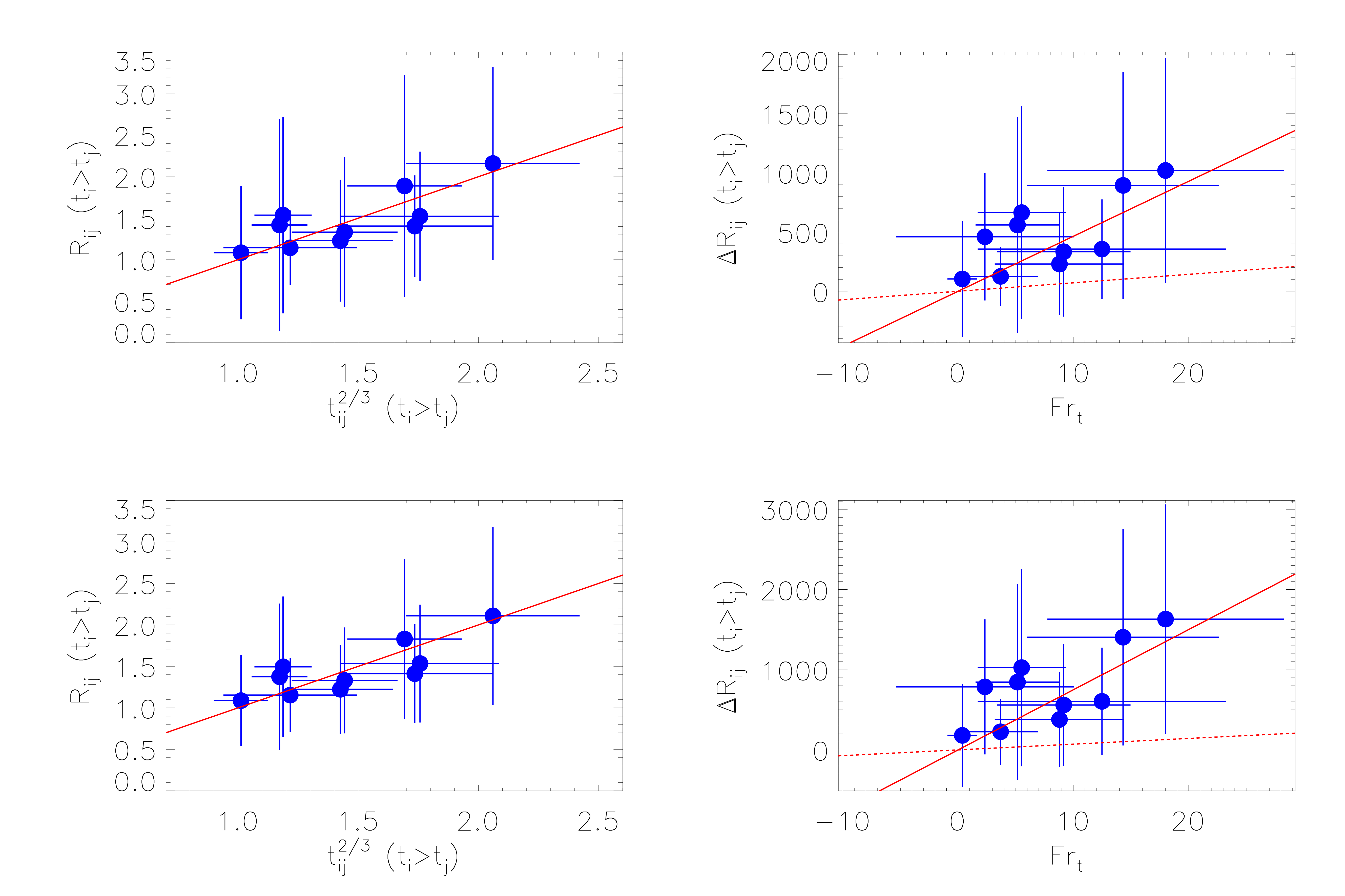}
\caption{Top left panel shows the dependence of $R_{ij}$ on $t_{ij}^{2/3}$. Solid red line shows $R_{ij} = t_{ij}^{2/3}$, based on the 
determined parameters by the elliptical accretion disk model. Top right panel shows the dependence of $\Delta R_{ij}$ on $Fr_t$, 
based on the determined parameters by the elliptical accretion disk model. In the top right panel, dashed red line and solid red line 
show the $Fr_t=\Delta R_{ij}\frac{(GM_{BH}\pi)^{2/3}}{2c^2}$ and $Fr_t=6.5\Delta R_{ij}\frac{(GM_{BH}\pi)^{2/3}}{2c^2}$, respectively. 
Bottom panels show the corresponding results based on the determined parameters by the circular accretion disk plus spiral 
arms model. In bottom left panel, solid red line shows $R_{ij} = t_{ij}^{2/3}$. In bottom right panel, dashed red line and solid red 
line show the $Fr_t=\Delta R_{ij}\frac{(GM_{BH}\pi)^{2/3}}{2c^2}$ and $Fr_t=10.5\Delta R_{ij}\frac{(GM_{BH}\pi)^{2/3}}{2c^2}$, 
respectively.}
\label{res}
\end{figure*}

\begin{figure*}
\centering\includegraphics[width = 18cm,height=8cm]{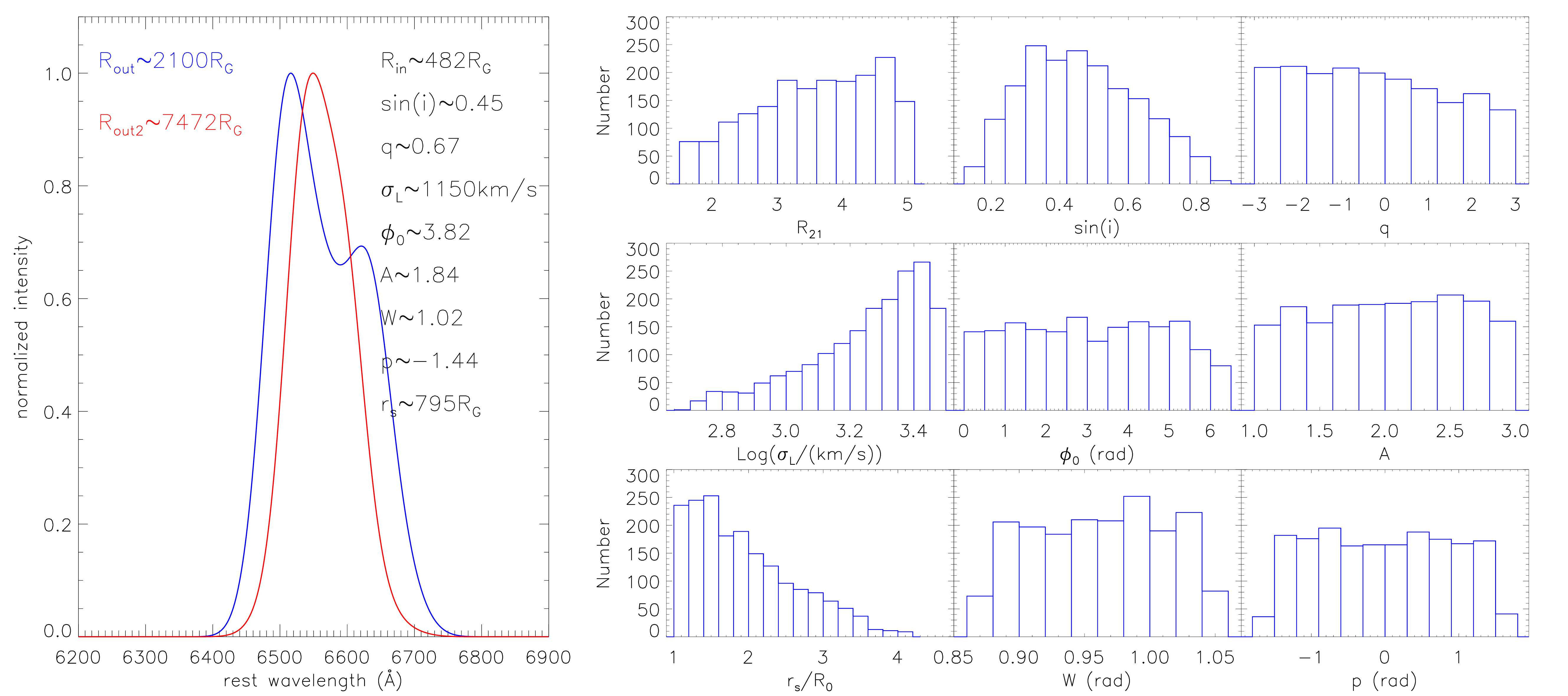}
\caption{Left panel shows an example on the broad H$\alpha$ with double-peaked profile (in blue) expected by the circular
accretion disk plus spiral arms model with smaller $R_{out}$, but with single-peaked profile (in red) expected by the model with
larger $R_{out2}$. The applied model parameters, except the $R_{out}$ and $R_{out2}$ marked in the top left corner, are marked in the
top right corner. The other panels show the distributions of the model parameters of the 1825 cases which have profiles changed from
double-peaked to single-peaked, through the circular accretion disk plus spiral arms model.}
\label{arm}
\end{figure*}

%%%2.3
	Based on the simulated results, Fig.~\ref{dis} shows the distributions of the parameters of $R_{21}=\frac{R_{out2}}{R_{out}}$, 
$\sin(i)$, $q$, $\sigma_L$, $e$ and $\phi_0$ of the 1977 cases. It is clear that the detected changed profile from double-peaked to 
single-peaked has not sensitively dependence on the model parameters. Moreover, an example of expected line profile variability is 
shown in Fig.~\ref{mod} by changing the model parameter of $R_{out}$ but with the other model parameters to be fixed, leading the model 
expected double-peaked profile changed to single-peaked profile due to small $R_{out}$ varied to large $R_{out2}$.

%%%2.4
	Before proceeding further, considering the changed line profile from double-peaked to single-peaked due to increases $R_{out}$, 
further clues could be obtained to support central TDE and/or to estimate locations of BLRs in TDE candidates. Accepted the time 
dependent evolution of $R_{TDE,out}$ (shown in equation 1 in the Introduction) discussed in \citet{gm14}, If there were clear time 
information for the double-peaked line profile (at time $t_d$) and the single-peaked line profile (at time $t_s$), then through the 
$t_d$ and $t_s$ and the determined $R_{out2}$, $R_{out}$, assumed the central disk-like BLRs have the similar expanded properties as 
the fallback TDE debris, we will have the following two sub-equations
\begin{equation}
\begin{split}
	&(\frac{t_s}{t_d})^{2/3}\sim\frac{R_{out2}}{R_{out}} \\
	2(\frac{GM_{BH}t_s^2}{\pi^2})^{1/3}&-2(\frac{GM_{BH}t_d^2}{\pi^2})^{1/3}\ge(R_{out2}-R_{out}) \\
\end{split}
\end{equation}
Therefore, the first sub-equation can be applied to test whether there are expected increased outer boundaries of the disk-like 
BLRs related to TDE debris. Meanwhile, considering $R_{out}$ in units of $R_G=\frac{GM_{BH}}{c^2}$ ($c$ as the light speed), the 
second sub-equation above can be re-written as
\begin{equation}
\begin{split}
	&Fr_t=t_s^{2/3}-t_d^{2/3}=k_s\Delta R_{out}\times\frac{(GM_{BH}\pi)^{2/3}}{2c^2} \ \ (k_s\ge1)\\
	&\Delta R_{out}=\frac{R_{out2}}{R_G}~-~\frac{R_{out}}{R_G}
\end{split}
\end{equation}
It is clear that the equation 3 can provide an independent method to estimate the locations of the disk-like BLRs in central 
accretion disk in an assumed TDE, if the central BH mass has been measured. For the equation 3, unless the disk-like BLRs and 
the fallback TDE debris have the same outer boundaries, the factor $k_s=1$ is preferred, otherwise $k_s~>~1$.

%%%2.5
	Based on the model dependent results, it is interesting to check whether double-peaked broad lines changed to single-peaked 
can be detected in a real TDE candidate. Among the reported TDE candidates, there is one TDE candidate AT 2018hyz of which broad 
H$\alpha$ shows double-peaked features in the early stage but single-peaked features in the late stage. The detailed discussions 
on both the photometric variability and the spectroscopic properties can be found in \citet{hf20}. Based on the collected data points 
binned with 4.5\AA~from the Fig.~3 in \citet{hf20} for the spectra after subtractions of host galaxy contributions, the clear line 
profiles of the five broad H$\alpha$ with $\Delta t\sim27, 51, 117, 120, 164$days ($\Delta~t$ as the time interval since the discovery 
at MJD=58432) are shown in the left panel of Fig~\ref{tde}. Here, we should note that the continuum emissions underneath the broad 
H$\alpha$ have been removed by a linear function determined by the data points with rest wavelength from 6200\AA~to 6350\AA~and from 
6820\AA~to 7000\AA. Meanwhile, due to broad H$\alpha$ not complete and/or due to no apparent broad features, the spectroscopic features 
around 6564\AA~ are not considered for the two spectra with $\Delta t\sim199,~364$days. Clearly, the first two broad H$\alpha$ with 
$\Delta t\sim27,~51$days have apparent double-peaked features. However, the other three broad H$\alpha$ with $\Delta t\sim117, 120, 
164$days have relatively smooth profiles without apparent double-peaked features.

%%%2.6
	Now, it is interesting to check whether considering the increased outer boundaries $R_{out}$ can be applied to describe the 
profile variability in the five broad H$\alpha$ in AT 2018hyz. Here, the elliptical accretion disk model are applied to simultaneously 
describe the five broad H$\alpha$, with the following model parameters. The five values of $R_{out}$ are increased with time for the 
five broad H$\alpha$. There are also five values of $\sigma_L$ for the five broad H$\alpha$. The varying $\sigma_L$ can be commonly 
accepted, considering the evolution of disk-like BLRs covering different emission regions. Besides the $R_{out}$ and $\sigma_L$, the 
other model parameters are the same for each broad H$\alpha$, when to describe the broad H$\alpha$. Therefore, when the model applied 
to describe the five broad H$\alpha$, there are 15 model parameters, $R_{in}$, five $R_{out}$, $\sin(i)$, $q$, $e$, five $\sigma_L$ 
and $\phi_0$. Then through the Levenberg-Marquardt least-squares minimization technique (the MPFIT package) \citep{mc09}, left panel 
of Fig.~\ref{tde} shows the best descriptions to the five broad H$\alpha$, with the corresponding $\chi^2/dof\sim1.05$ ($dof$ as degree 
of freedom). The determined model parameters are listed in the fourth column in Table~1. The five values of $R_{out}$ increased with 
time are also marked in the left panel of Fig.~\ref{tde}.

%%%2.7
	As discussed in \citet{hf20} through applications of theoretical TDE model \citet{gr13, gm14, mg19} to describe long-term 
photometric variability of AT 2018hyz, the time interval is about $-43_{-9}^{+8}$days between the starting time $t_0$ for the assumed 
TDE and the time for the discovery of AT 2018hyz. Therefore, the corresponding time information in rest frame for the five observed 
broad H$\alpha$ since $t_0$ are about $t_1 = (27+43)_{-9}^{+8}$days, $t_2 = (51+43)_{-9}^{+8}$days, $t_3 = (117+43)_{-9}^{+8}$days, 
$t_4 = (120+43)_{-9}^{+8}$days and $t_5=(164+43)_{-9}^{+8}$days. Then, for any two of the broad H$\alpha$, there are ten $R_{out}$ 
ratios, $R_{ij}=R_{out,t_i}/R_{out,t_j}$ ($R_{out,t_i}$ as the determined $R_{out}$ for the broad H$\alpha$ at $t_i$) with $t_i>t_j$, 
and also corresponding ten time ratios, $t_{ij}=t_i/t_j$ with $t_i>t_j$. 

%%%2.8
	Top left panel of Fig.~\ref{res} shows the dependence of the ratio $R_{ij}$ on the ratio $t_{ij}^{2/3}$, leading to a linear 
dependence as expected by the first sub-equation in equation 2. There are only ten data points, therefore, there are no further 
discussions on the robust of the linear dependence but to show the determined Spearman Rank correlation coefficient about 0.58 
($P_{null}\sim0.08$). Furthermore, top right panel of Fig.~\ref{res} shows the dependence of the $R_{out}$ difference 
$\Delta R_{ij}=R_{out,t_i}-R_{out,t_j}$ ($t_i>t_j$) on the corresponding $Fr_t$ with Spearman Rank correlation coefficient about 0.62 
($P_{null}\sim0.05$). Accepted the central BH mass $3.5\times10^6{\rm M_\odot}$ in AT 2018hyz as discussed in \citet{hf20}, $k_s\sim6.5$ 
applied in equation 3 can describe the shown dependence of $\Delta R_{ij}$ on $Fr_t$. Therefore, the outer boundaries of the disk-BLRs 
for the broad H$\alpha$ are about one sixth of the outer boundaries of the fallback TDEs debris in AT 2018hyz.

%%%2.9
	Before ending the section, three additional points are noted. For the first point, it is necessary to check whether disk 
rotating have strong effects on the profile variability of broad H$\alpha$ in AT 2018hyz. Based on the determined eccentricity, 
inner and outer boundaries of the disk-like BLRs, the expected disk precession period as discussed in \citet{se03} is about $T_{pre}$
\begin{equation}
T_{pre}\sim 10.4M_{BH,6}\frac{1+e}{(1-e)^{3/2}}R_{em,3}^{2.5}{\rm yrs}
\end{equation}
with $M_{BH,6}$ as the BH mass in units of $10^6{\rm M_\odot}$ and $R_{em,3}$ as the distance in units of $10^3R_{G}$ of disk-like 
BLRs to central BH and $e$ as the eccentricity of the disk-like BLRs. For AT 2018hyz, $M_{BH,6}$ is about 3.5, $e\sim0.133$, and 
the $R_{em,3}$ can be estimated to be 0.612 by the flux-weighted distance from central BH after considering the model parameters 
in the elliptical accretion disk model to describe the broad H$\alpha$ with $\Delta t=17$days. Therefore, the expected disk 
precession period is about 15years, very longer than the time intervals only around 150days for the broad H$\alpha$ in AT 2018hyz. 
Moreover considering the model parameters for the other broad H$\alpha$, due to larger outer boundary, longer procession period 
than 15years can be expected. Therefore, there are few effects of disk procession on profile variability of broad H$\alpha$ in 
AT 2018hyz. For the second point, the listed model parameters are not similar as the ones reported in \citet{hf20}, mainly due 
to no considerations of extra Gaussian components in the broad H$\alpha$ in this manuscript. Therefore, there are no further 
discussions on different model parameters in this manuscript and in \citet{hf20}. For the third point, the determined $\sigma_L$ 
at $\Delta t=51$days is very different from the $\sigma_L$ at the other epochs, probably due to different local temperatures. 
As the shown Fig.~8 in \citet{mg19}, after the peak, the temperature decreases slightly near peak and then gradually increases as 
the luminosity decreases for common TDE candidates. Therefore, in AT 2018hyz, the expected temperature at $\Delta t=51$days should 
be the minimum value, leading to the smaller temperature dependent $\sigma_L$.

\begin{table}
\caption{Model parameters for the circular accretion disk plus spiral arms model}
\begin{tabular}{llll}
\hline\hline
        par & units & model & Arm  \\
\hline\hline
        $R_{in}$ & $\rm R_G$ & [100,~600]    &  268$\pm$170\\
        $R_{out}$ & $\rm R_G$ & $k_0\times R_{in}$ ($k_0\in[1.5,~5]$) & 1469$\pm$266 \\
        $R_{out2}$ & $\rm R_G$ & $k\times R_{out}$ ($k\in[1.5,~5]$) &   1695$\pm$330 \\
                   & $\rm R_G$ &                                     &  2074$\pm$470 \\
                   & $\rm R_G$ &                                     &  2255$\pm$630 \\
                   & $\rm R_G$ &                                     &  3100$\pm$950 \\
        $\sin(i)$ &   & [0.3, ~0.9]   & 0.69$\pm$0.05 \\
        $q$ &         & [-3, ~3]  &   -0.267$\pm$0.395\\
        $\sigma_L$ & km/s & [500, ~3000]  &  1800$\pm$165\\
        &     km/s    &            &     700$\pm$100 \\
        &     km/s    &            &     1788$\pm$235 \\
        &     km/s    &            &     1877$\pm$207 \\
        &     km/s    &            &     3000$\pm$337 \\
        $\phi_0$      & rad  & [0, ~2$\pi$]  &  1.87$\pm$0.63 \\
	$A$     &     &   [1, ~~3]  & 2.72$\pm$0.42 \\
	$W$     & degree & [10, ~~60]  & 58$\pm$27 \\
	$p$     & degree & [-90, ~~90] & 33$\pm$10 \\
	$r_s$ & $R_G$   & [$r_0$, ~~0.8$r_1$] & 270$\pm$160 \\
\hline\hline
\end{tabular}\\
Notice: The first column and the second column show the applied model parameters and the corresponding units. The third column 
shows the limited range ([lower boundary, upper boundary]) of each model parameter in the elliptical accretion disk model leading to the
artificial $f_{\lambda1}$ and $f_{\lambda2}$. The fourth column shows the determined model parameters in the circular accretion disk plus 
spiral arms model applied to describe the time dependent line profiles of broad H$\alpha$ in AT 2018hyz. In the fourth column, the five 
values in $R_{out}$ and $R_{out2}$ and the five values in $\sigma_L$ are for the five broad H$\alpha$ with $\Delta t=17, 51, 117, 
120, 164$days.
\end{table}

	Before ending the section, there is one point we should note. In the accretion disk models above for the simulations, the 
corresponding boundaries of emission regions are in units of $R_G$, leading the simulations not to depend on BH masses or boundaries 
of BLRs in physical distance units, indicating the simulations can be commonly applied in BLRs related to any TDEs. In other words, 
the simulations are through the known accretion disk models, previously given BH masses and/or boundaries of BLRs have no effects 
on the corresponding simulation results, only except the $r_{in}$ and $r_{out}$ described in physical distance units of 
light-days and/or pc and/or km. In one word, through the simulations by accretion disk models with or without given values 
of BH mass and/or boundaries of BLRs, the same results can be determined. Therefore, there are no discussions on effects of given 
BH mass and/or boundaries of BLRs in the manuscript.

\section{Another explanations to the changed profiles of double-peaked broad emission lines}

	Considering the accreting fallback TDEs debris commonly in elliptical orbits as discussed in \citet{gm14}, the elliptical 
accretion disk model is mainly considered and discussed above. However, there are several another explanations to the changed 
profiles of double-peaked emission lines related to disk-like BLRs. In the section, the following two additional explanations are 
mainly discussed.

	Considering eccentricity of the disk-like BLRs to be zero, the improved circular disk plus spiral arms model \citep{se03} is 
discussed with ten model parameters. Besides the model parameters (with eccentricity to be zero) applied in the elliptical accretion 
disk model \citep{el95}, there are four addition parameters, the contrast ratio $A$ for the arms relative to the rest of the disk, the 
width $W$ and pitch angle $p$ for the arms, and the starting radius $r_{s}$ (in units of $R_{G}$) of the arms. As an optional model to 
explain double-peaked broad lines, through dynamical properties of disk-like BLRs. It is obvious that larger outer boundary of the 
disk-like BLRs in the circular disk plus spiral arms model can also lead to smaller peak separation of double-peaked broad emission 
lines.

	Similar as what we have done in Section 2 on the elliptical accretion disk model, the following procedure is applied based on the 
circular accretion disk plus spiral arms model. First, the ten model parameters are randomly collected within the limited ranges listed 
in Table~2, leading to a model created line profile $f_{\lambda1}$. Second, a new value of the outer boundary $R_{out2}$ is randomly 
collected by $R_{out2}\sim k\times R_{out}$, leading to the new model created line profile $f_{\lambda2}$ with different outer boundaries 
but with the same values for the other model parameters. Third, if $f_{\lambda1}$ has more than two peaks but $f_{\lambda2}$ has only 
one peak, we accept the case is the one which has the changed profile from double-peaked to single-peaked through the circular accretion 
disk plus spiral arms model. Then, to repeat the three above steps 50000 times, there are 1825 cases having the changed profiles from 
double-peaked to single-peaked, indicating the probability roughly about 3.65\% (1825/50000) to detect changed profiles in multi-epoch 
optical spectrum with considerations of the evolved outer boundaries of the disk-like BLRs in TDE candidates. Left panel of Fig.~\ref{arm} 
shows an example on the broad H$\alpha$ with double-peaked profile expected by the circular accretion disk plus spiral arms model with 
smaller $R_{out}$, but with single-peaked profile expected by the model with larger $R_{out2}$. The other panels of Fig.~\ref{arm} show 
the distributions of the model parameters of the 1825 cases. Therefore, the circular accretion disk plus spiral arms model can also lead 
to broad line profiles changed from double-peaked to single-peaked.

	Then, the circular accretion disk plus spiral arms model has been also applied to describe the variability of broad H$\alpha$ 
in AT 2018hyz through the Levenberg-Marquardt least-squares minimization technique, the determined model parameters are listed in 
the last column of Table~2, and the best fitting results are shown in the right panel of Fig.~\ref{tde} with the corresponding 
$\chi^2/dof\sim1.02$. And then, based on the determined model parameters by the circular accretion disk plus spiral arms model, bottom 
panels of Fig.~\ref{res} show the corresponding dependence of the ratio $R_{ij}$ on the ratio $t_{ij}^{2/3}$, and the dependence of 
the $R_{out}$ difference $\Delta R_{ij}=R_{out,t_i}-R_{out,t_j}$ ($t_i>t_j$) on the corresponding $Fr_t$. Strong linear correlations 
can be found in the bottom panels, with the determined Spearman Rank correlation coefficients about 0.62 ($P_{null}\sim0.02$) and 0.72 
($P_{null}\sim0.05$) for the results shown in the bottom left panel and in the bottom right panel, respectively. Meanwhile, accepted 
the central BH mass $3.5\times10^6{\rm M_\odot}$ in AT 2018hyz, $k_s\sim10.5$ applied in equation 3 can describe the shown dependence 
of $\Delta R_{ij}$ on $Fr_t$ in the bottom right panel of Fig.~\ref{res}. Therefore, assumed the circular accretion disk plus spiral 
arms model preferred in AT 2018hyz, the outer boundaries of the disk-BLRs for the broad H$\alpha$ are about one tenth of the outer 
boundaries of the fallback TDEs debris in AT 2018hyz.

	Unfortunately, there is not enough evidence to confirm which disk model, the elliptical accretion disk model or the circular 
accretion disk plus arms model, is preferred in the TDE candidate AT 2018hyz. However, considering that the elliptical orbitals are 
common in accreting fallback TDEs debris, the results through the elliptical accretion disk model are preferred, but the results through 
the circular accretion disk plus arms model can be accepted as potentially supplementary results.

	Besides the evolution of outer boundary of disk-like BLRs expected by evolution of TDEs debris, disk instability (such as 
sudden hot spots, and/or radiation driven radial outflows) could be also applied to describe the broad line profiles changed from 
double-peaked to single-peaked, such as the more recent work on line profile variability of broad Balmer emission lines in 
NGC 1566 in \citet{om24} due to strong scale-height-dependent turbulence. In other words, double-peaked broad line coming from 
disk-like BLRs can be changed to single-peaked broad line, after considering an addition emission component related to disk 
instability at one epoch. As discussed and shown in \citet{hf20}, besides the double-peaked emission component in broad H$\alpha$, 
the extra Gaussian component can be accepted as the additional emission component related to disk instability. The extra Gaussian 
component has its strength comparable to the strength of the double-peaked emission component, see results in Fig.~6 and 
Fig.~8 in \citet{hf20}, indicating that the emission strength of the extra emission component related to the disk instability 
should be comparable to the common disk emission strength. In other words, the emissions from the process related to the disk 
instability should lead to apparent sudden bursts at the epochs for the disk instability, leading to expected apparent sudden 
flares in the optical light curves. However, after checking the photometric light curves in AT 2018hyz (see Fig,~1 in \citealt{hf20}), 
the light curves are smooth enough. Therefore, the scenario on disk instability is not preferred to explain the changed profiles 
in broad H$\alpha$ in AT 2018hyz. Probably, in the near future, in one TDE with apparent sudden bursts/flares in optical light 
curves, strong turbulence related to disk instability could have contributions to explain broad emission lines having changed 
line profiles.

	Besides the evolution of outer boundary of disk-like BLRs expected by evolution of TDEs debris and the probable disk 
instability, fountain-like AGN feedback proposed in \citet{wk23} can also be applied to explain the line profile variability 
of broad emission lines. However, our main results are discussed in the physical framework of TDEs with time scales around tens 
to hundreds of days, which is very smaller than the time scale about 10 (or more) years for fountain-like AGN feedback leading 
to line profile variability of broad emission lines. Therefore, there are no further discussions on the proposed scenario 
in \citet{wk23} in our manuscript. Furthermore, as discussed in \citet{om24}, a low-optical-depth wind could lead to 
drift of broad line profiles. Therefore, considering the proposed scenarios in \citet{wk23, om24} on contributions of both 
wind and turbulence, not only changed profile from double-peaked to single-peaked but drifted line profile can be expected 
through the evolutions of broad emission lines. In the near future, to detect such unique variability properties of broad 
line profiles will provide interesting clues to support existence of turbulence and/or wind in TDEs with un-smooth light 
curves. Meanwhile, variability in outer radius of BLRs could also be due to variability of ionization luminosity, as expected 
by the reverberation mapping technique determined R-L relation between radius of BLRs and central continuum luminosity 
\citep{kas00, bd13}. However, considering the smooth decline trend in the AT 2018hyz, central continuum luminosity weaken with 
time should lead outer radius of the disk-like BLRs to decrease over time, indicating more apparent double-peaked features. 
Therefore, the variability of outer radius by R-L empirical relation is not preferred in the AT 2018hyz. In the near future, 
to determine detailed time dependent spatial structures of disk-like BLRs through the known GRAVITY interferometric technique 
\citep{gs18, ga21, ga24} in such TDEs with changed profiles of broad emission lines will provide further clues to support 
or to be against our proposed scenario in the manuscript.

\section{Summary and Conclusions}

	The final summary and conclusions are as follows. 
\begin{itemize}	
\item Not similar as rotations in disk-like BLRs lying into central accretion disk leading to profile variability in broad line AGN, 
	evolutions of outer boundaries of disk-like BLRs related to TDE debris can also lead profile variability of broad emission 
	lines, especially lead to changed profile from double-peaked to single-peaked. 
\item Based on simulated results by standard elliptical accretion disk model preferred after considering elliptical orbitals 
	for accreting fallback TDEs debris, the probability is about 3.95\% to detected profile changed from double-peaked to 
	single-peaked in multi-epoch broad emission lines.
\item Double-peaked broad H$\alpha$ in early stages but single-peaked broad H$\alpha$ can be detected in the TDE candidate AT 2018hyz.
\item Only considering increased outer boundaries of the disk-like BLRs can lead to well accepted descriptions to profiles of the 
	multi-epoch broad H$\alpha$ in AT 2018hyz.
\item The dependence of outer boundary ratio $R_{ij}$ on time interval ratio $t_{ij}$ is well consistent with expected results by the 
	evolution properties of fallback TDE debris in AT 2018hyz.
\item Through the dependence of outer boundary difference $\Delta R_{ij}$ on evolution time ratio $Fr_t$, outer boundary of the 
	fallback TDE debris is about 6.5times of the outer boundary of the disk-like BLRs lying into central accretion disk in AT 2018hyz.
\item Although the elliptical accretion disk model is preferred for the broad H$\alpha$ in the TDE AT 2018hyz, results on the 
	improved circular accretion disk plus spiral arms model are also checked and discussed, indicating the outer boundary of 
	the fallback TDE debris is about 10.5times of the outer boundary of the disk-like BLRs lying into central accretion disk in 
	AT 2018hyz.
\item Beside the accretion disk origin for the broad H$\alpha$, disk instability can also be applied to explain the changed 
	profiles of broad emission lines. However, applications of disk instability can lead to expected optical flares in photometric 
	light curves. After considering the smooth photometric light curves in AT 2018hyz, the disk instability is not preferred to 
	explain the changed profiles of broad H$\alpha$ in AT 2018hyz.
\item Such unique profile variability in broad lines from double-peaked to single-peaked could be accepted as further clues to 
	support a central TDE, besides the commonly applied photometric variability properties.
\end{itemize}

%\section*{Acknowledgements}
\begin{acknowledgements}
Zhang gratefully acknowledge the anonymous referee for giving us constructive comments and suggestions to greatly 
improve the paper. Zhang gratefully thanks the kind financial support from GuangXi University and the kind grant support from 
NSFC-12173020 and NSFC-12373014, and the support from Guangxi Talent Programme (Highland of Innovation Talents). This manuscript 
has made use of the MPFIT package (\url{http://cow.physics.wisc.edu/~craigm/idl/idl.html}).
\end{acknowledgements}

%\section*{Data Availability}
%The data underlying this article will be shared on reasonable request to the corresponding author
%(\href{mailto:xgzhang@gxu.edu.cn}{xgzhang@gxu.edu.cn}).

\label{lastpage}
\end{document}